\documentstyle[aps,epsfig,pre]{revtex}
\begin{document}
\title{Urn model of separation of sand}
\author{Adam  Lipowski$^{1),2)}$ and Michel
Droz$^{1)}$}
\address{$^{1)}$ Department of Physics, University of Gen\`eve, CH 1211
Gen\`eve 4, Switzerland\\
$^{2)}$ Department of Physics, A. Mickiewicz 
University, 61-614 Poznan, Poland}
\date{\today}
\maketitle
%%%%%%%%%%%%%%%%%%%%%%%%%%%%%%%%%%%%%%%%%%%%%%%%%%%%%%%%%%%%%%%%%
\begin{abstract}
We introduce an urn model which describes spatial separation of sand.
In this dynamical model, in a certain range of parameters spontaneous 
symmetry breaking takes place and equipartitioning of sand into two 
compartments is broken.
The steady-state equation for an order parameter, a critical line, and the 
tricritical point on the phase diagram are found exactly.
Master equation and the first-passage problem for the model are solved 
numerically and the results are used to locate first-order transitions.
Exponential divergence of a certain characteristic time shows
that the model can also exhibit very strong metastability.
In certain cases characteristic time diverges as $N^{z}$, where $N$ 
is the number of balls and $z=\frac{1}{2}$ (critical line), $\frac{2}{3}$ 
(tricritical point), or $\frac{1}{3}$ (limits of stability).
\end{abstract}
\pacs{}
%\begin{multicols}{2}
\tightenlines
%%%%%%%%%%%%%%%%%%%%%%%%%%%%%%%%%%%%%%%%%%%%%%%%%%%%%%%%%%%%%%%%%%%%
%%%%%%%%%% 
\section{introduction}
Recently, granular systems are intensively studied using methods of statistical
physics.
The main motivation of this research is to get basic understanding of these 
technologically important materials whose behaviour, however, very often 
appears elusive and  mysterious~\cite{SWINNEY,MUZZIO}.

One of the classical experiments in this field concerns the spatial separation
of shaken sand~\cite{SCHLICH}.
In this experiment one uses a box which is divided into two equal
compartments
by a wall which has a narrow horizontal slit at a certain height.
When the box is filled with a sand and shaken one observes that under certain
conditions (e.g., frequency and amplitude of shaking) one compartment of a 
box is preferentially filled with sand. 
Theoretical arguments to understand how the symmetry breaking arises in this 
experiment, were given by Eggers~\cite{EGGERS}.
In his approach, Eggers studied a continuous model based on partial differential 
equations which takes into account
certain particular characteristics of shaken granular systems~\cite{KUMARAN}.
Namely the fact that effective temperature of a granular system decreases when
the density of particles increases.
This at first counterintuitive feature of granular systems is now understood as
a result of inelastic collisions between particles which effectively cool the 
system~\cite{ZANETTI}.
This feature is also responsible for the symmetry breaking in the experiment 
with shaken sand.
Indeed, with nonequal distribution of sand the temperature in the compartment
with majority of particles is substantially reduced.
Consequently, particles from this compartment are less likely to
cross the slit.
Eggers has shown that a continuous phase transition accompanies the 
symmetry breaking.
The approach used by Eggers was generalized for a larger number of compartments
separating the box.
It was shown that in such a case the phase transition becomes 
discontinuous~\cite{MEER}.

Eggers's model is however quite complicated and its analysis must be 
supplemented with some simplifying assumptions.
The analysis of fluctuations or dynamical properties seems to be
particularly difficult for this model.
A possible alternative might be to examine a different, presumably simpler
model, which would nevertheless grasp the essence of the phenomena.

The main goal of the present paper is to define and analyze such a model.
Our model is a certain generalization of the urn model introduced by 
Ehrenfest~\cite{EHREN}.
In Ehrenfest model one has balls which are distributed between two urns and
certain dynamical rules according to which balls can change their location.
Ehrenfest model greatly contributed to the understanding of some basic notions 
of statistical mechanics like the equilibrium or approach to the equilibrium.
Various modifications of this model were examined~\cite{MODIF}.

At first sight it seems natural that Ehrenfest model could provide an
approximate description of experiments with shaken sand.
However, none of its generalizations described in the literature exhibits
a spontaneous symmetry breaking~\cite{DES}.
The modification which we propose does exhibit a spontaneous 
symmetry breaking.
Although the model is rather simple it exhibits a rich behaviour.
In addition to having a line of continuous transitions, 
the model has a tricritical point and a line of discontinuous transitions. 
The discontinuous transitions are screened by very strong metastability.
Existence of this, unobservable in fact (due to metastability) transition in 
our dynamical model requires, however, some comments.
In equilibrium systems it is the free energy which unambiguously determines
the phase which is stable for a given set of parameters.
For our dynamical model we do not know how to calculate the free energy.
Nevertheless, certain features enabled us to locate the discontinuous 
transitions in our model.
Namely probability distribution of the order parameter exhibits characteristics 
which are typical to equilibrium systems with discontinuous transitions.
Additional evidence of such a transition in our model comes form the 
analysis of 
characteristic times, which we determine solving numerically the
first-passage problem.

Our paper is organized as follows.
In section II we introduce the model and examine its steady-state properties.
In section III, from the solution of the master equation we analyze the 
steady-state probability distribution and fluctuations in the model.
Section IV is devoted to the metastable properties and characteristic times 
obtained from the solution of the first-passage problem.
Conclusions and possible extensions of our work are discussed in section V.
%%%%%%%%%%%%%%%%%%%%%%%%%%%%%%%%%%%%%%%%%%%%%%%%%%%%%%%%%%
\section{Model and its steady state}
Our model is defined as follows:
$N$ particles are distributed between two urns A and B and the number of
particles in each urn is denoted as $N_A$ and $N_B$, respectively 
($N_A+N_B=N$).
Particles in a given urn are subject to thermal fluctuations and the 
temperature $T$ of the urn depends on the number of particles in it.
To mimic the effect that a larger number of particles leads to a smaller 
temperature, we use the following rule:
\begin{equation}
T(n_{A,B})=T_0+\Delta(1-n_{A,B}),
\label{temp}
\end{equation}
where $n_{A,B}=N_{A,B}/N$ is a fraction of a total number of particles in a 
given urn and $T_0$ and $\Delta$ are positive constants.
With such a choice, temperatures of urns changes linearly with $n_{A,B}$ from 
the value $T_0$ to $T_0+\Delta$.
For granular systems the relation between temperature and the number of
particles is very complicated and depends on certain parameters like density 
of particles or the type of driving~\cite{KUMARAN}.
For the experiment with separation of sand certain arguments suggest that 
$T\sim n_{A,B}^{-2}$ would be more suitable~\cite{EGGERS}.
However, with such a choice the location of the critical point changes with 
$N$.
As expected on general grounds, the observed behaviour for finite $N$ is 
only a crossover and a well-defined phase transition exists only in the limit
$N\rightarrow\infty$.
Our choice (\ref{temp}), leads to results which are well-defined in the 
thermodynamic limit $N\rightarrow\infty$.
However, it would be interesting to examine other choices too.

Next, we assume that particles within a given urn obey the standard 
Maxwell-Boltzmann distribution.
It means that their distribution changes with the height $z$ (above the 
bottom of an urn) as 
\begin{equation}
p_{A,B}(z)=\frac{mgN_{A,B}}{T(n_{A,B})}{\rm exp}[\frac{-mgz}{T(n_{A,B})}],
\label{height}
\end{equation}
where $m$ is the mass of a particle and $g$ is the Earth acceleration.
One can easily calculate that the fraction of particles obeying
distribution (\ref{height}), which are above a certain height $h$ is given as
${\rm exp}[\frac{-mgh}{T(n_{A,B})}]$.
Next, we define dynamics of our model as follows:\\
(i) One of the $N$ particles is selected randomly.\\
(ii) With probability ${\rm exp}[\frac{-mgh}{T(n_{A,B})}]$ the selected 
particle changes urns.
($n_{A,B}$ is the fraction of particles present in the urn of the selected 
particle).\\
In the following we assume that temperature is measured in units 
such that $mgh=1$.
It means that our model is parametrized only by $T_0$ and $\Delta$.
The Ehrenfest model is recovered in the limit of infinite temperature 
where every selected particle changes urns.
 
\begin{figure}
\centerline{\epsfxsize=9cm 
\epsfbox{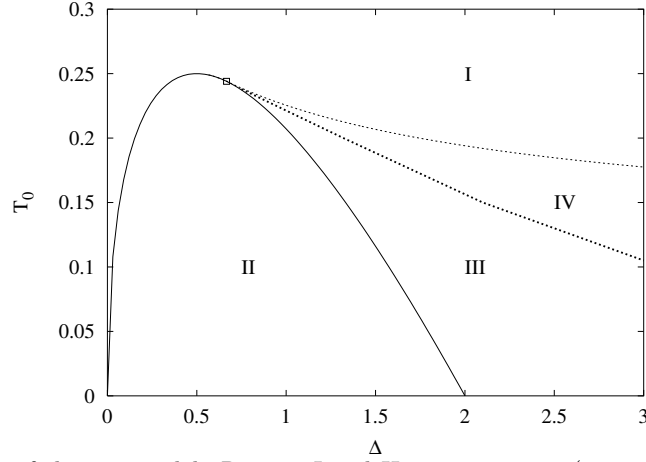}}
%\figspace
\caption{
The phase diagram of the urn model.
Regions I and II are symmetric ($<\epsilon>=0$) and asymmetric 
($<\epsilon>\neq 0$) phases, respectively.
In region III (IV) the symmetric (asymmetric) phase is metastable (see text).
The solid line is the critical line (\ref{critical}).
The thick dotted line is the line of first-order phase transitions.
The tricritical point ($\frac{2}{3},\frac{\sqrt 3 -1}{3}$) is denoted as $\Box$.
}
\label{diagram}
\end{figure}

\begin{figure}
\centerline{\epsfxsize=9cm 
\epsfbox{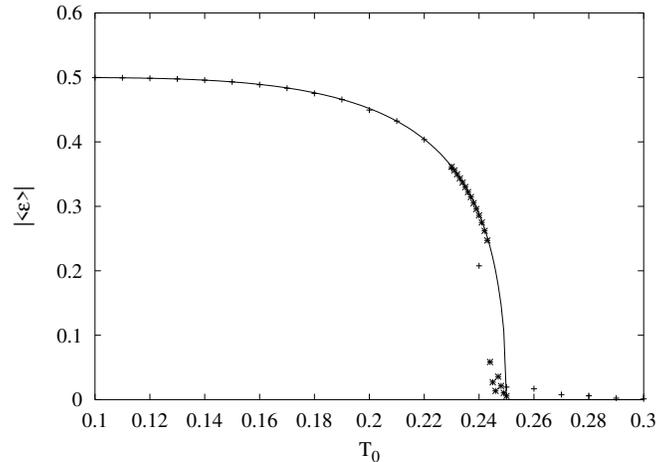}}
%\figspace
\caption{
The absolute value of the order parameter $|<\epsilon>|$ as a function of $T_0$
for $\Delta=0.5$.
Solid line corresponds to the numerical solution of eq.~(\ref{order}).
Monte Carlo simulations were made for $N=500$ (+) and 5000 ($\star$).
}
\label{forder}
\end{figure}

We define $\epsilon$ as a measure of the difference in the occupancy of the 
urns
\begin{equation}
\epsilon=\frac{N_A-N_B}{2N}.
\label{order} 
\end{equation}
For our model, one can find a simple, but exact equation which determines the 
time average $<\epsilon>$ in the steady-state.
Indeed, in the steady state the flux of particles changing their positions
from A to B equals to the flux from B to A.
It is easy to realize that since the selected particles are uncorrelated, 
the above requirement can be written as:
\begin{equation}
<N_A>{\rm exp}[\frac{-1}{T(<N_A/N>)}]=<N_B>{\rm exp}[\frac{-1}{T(<N_B/N>)}].
\label{flux} 
\end{equation}
Using the parameter $<\epsilon>$ and the relation $N_A+N_B=N$, 
eq.~(\ref{flux}) can be written as
%\begin{widetext}
\begin{equation}
(\frac{1}{2}+<\epsilon>){\rm exp}[\frac{-1}{T(\frac{1}{2}+<\epsilon>)}]=
(\frac{1}{2}-<\epsilon>){\rm exp}[\frac{-1}{T(\frac{1}{2}-<\epsilon>)}].
\label{flux1} 
\end{equation}
%\end{widetext}
Of course, eq.~(\ref{flux1}) admits symmetric solution $<\epsilon>=0$ for any
$T_0$ and $\Delta$.
However, a physically satisfactory solution must be stable with respect to 
fluctuations.
To make the stability analysis of $\epsilon=0$ solution of eq.~(\ref{flux1}) 
we have to expand it in powers of $\epsilon$.
Equating linear terms we obtain that the critical 
line which separates stable and unstable regions for symmetric solution is 
given as:
\begin{equation}
T_0^c=\sqrt{\Delta/2} -\Delta/2
\label{critical}
\end{equation}
This line is plotted as a solid line in Fig.~\ref{diagram}.
%%%%%%%%%%%%%%%%%%%%%%%%%%%%%%%%%%%%%%
%%%%%%%%%%%%%%%%%%%%%%%%%%%%%%%%%%%%%
What is more interesting, however, eq.~(\ref{flux1}) has also asymmetric 
solutions with $<\epsilon>\neq 0$.
Although such solutions cannot be written in a closed form, they can be
easily determined numerically.
An example of such a solution is shown in Fig.~\ref{forder}
In addition, expanding eq.~(\ref{flux1}) in powers of $<\epsilon>$, one can 
easily check that in the vicinity of the critical point 
$|<\epsilon>|\sim(T_0-T_0^c)^{1/2}$, which recovers an equilibrium mean-field 
exponent $\beta=1/2$ (we assume that $\Delta$ is kept const).
%%%%%%%%%%%%%%%%%%%%%%%%%%%%%%%%%%%%%%%
%%%%%%%%%%%%%%%%%%%%%%%%%%%%%%%%%%%%%%%%%%%
It turns out, however, that in a certain region of phase diagram both 
symmetric and asymmetric solutions are stable (see Fig.~\ref{diagram}.
It suggests that in that region a line of discontinuous transition should
exist.
This line would meet a critical line at a certain point which, by analogy with
equilibrium systems~\cite{BLUME}, can be called a tricritical point.
Location of this point can be easily determined from eq.~(\ref{flux1}).
Indeed, requiring that the third order (in $\epsilon$) coefficients are 
the same~\cite{even} and taking in account eq.~(\ref{critical}), we obtain the the tricritical
point is located at $\Delta=\frac{2}{3}, T_0=\frac{\sqrt 3 -1}{3}$.
However, it is by no means obvious how to determine the line of discontinuous
transitions in our model.
In equilibrium systems, one can calculate the free energy of each phase and
this line is determined from the condition that the free energy of both phases 
are equal.
Our model is not an equilibrium system and this procedure cannot be used.
Nevertheless, as we will show below, behaviour of certain quantities shows
striking similarity to equilibrium counterparts.
With this observation we will be able to locate discontinuous  transitions 
in our model.
To address this problem, we have to go beyond
the steady-state eq.~(\ref{flux1}), which is the subject of the next section.
%%%%%%%%%%%%%%%%%%%%%%%%%%%%%%%%%%%%%%%%%%%%%%%%%%%%%%%%%%%%%%%%%%%%%%%%%%%%
\section{master equation}
\begin{figure}
\centerline{\epsfxsize=6cm 
\epsfbox{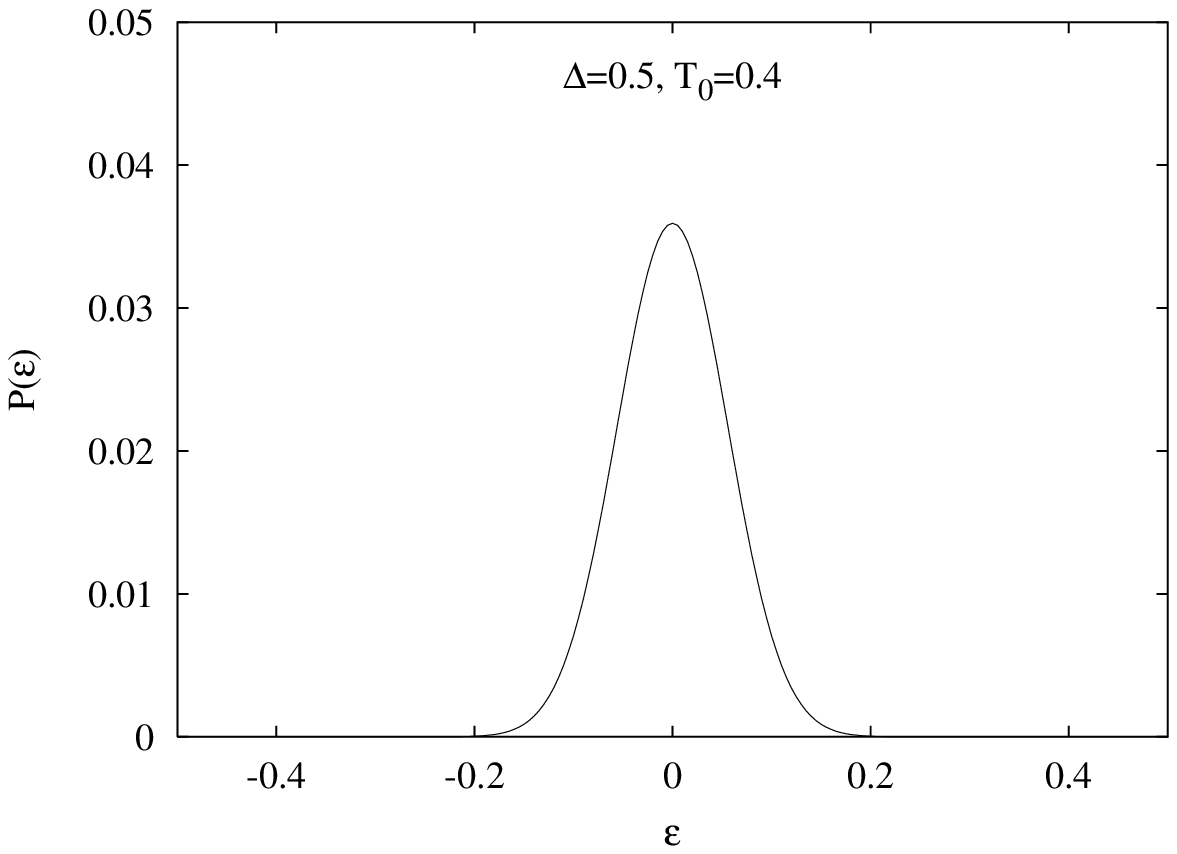}}
\vspace{-0.0cm}
\centerline{\epsfxsize=6cm 
\epsfbox{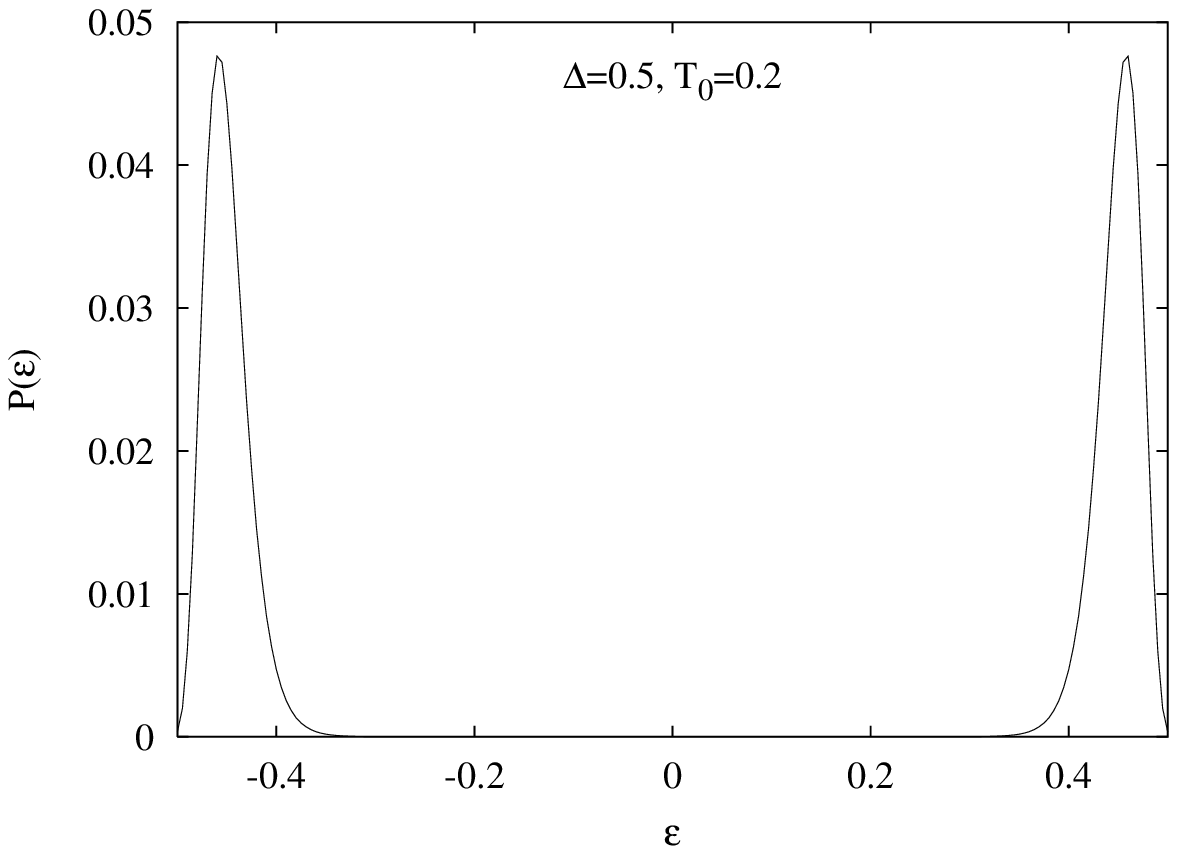}}
\centerline{\epsfxsize=6cm 
\epsfbox{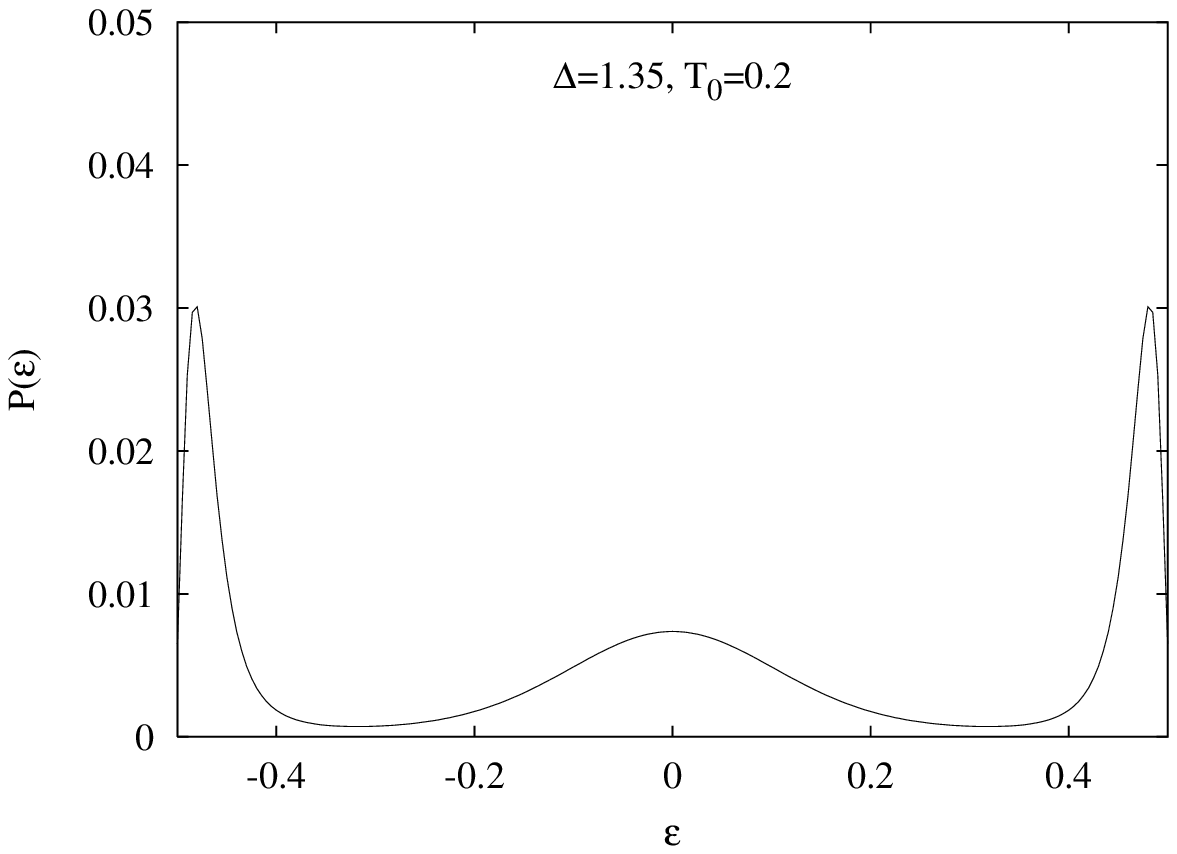}}
%\figspace
\caption{Probability distribution $p[N(\epsilon+\frac{1}{2}),t]$ obtained 
as a numerical solution of eq.~(\ref{evol1}) in the long time limit ($N=200$).
}
\label{config}
\end{figure}

For our urn model one can write  evolution equations for the probability 
distribution $p(M,t)$ that in a given urn (say A) at the time $t$ there are
$M$ particles.
These equations easily follow from the dynamical rules:
\begin{eqnarray}
p(M,t+1)=&&\frac{N-M+1}{N}p(M-1,t)\omega(N-M+1)+
\frac{M+1}{N}p(M+1,t)\omega(M+1)+\nonumber\\
&&p(M,t)\{\frac{M}{N}[1-\omega(M)]+\frac{N-M}{N}[1-\omega(N-M)]\}
 \ {\rm for}\ M=1,2\ldots N-1\nonumber\\
 p(0,t+1)=&&\frac{1}{N}p(1,t)\omega(1)+p(0,t)[1-\omega(N)],\nonumber\\
 p(N,t+1)=&&\frac{1}{N}p(N-1,t)\omega(1)+p(N,t)[1-\omega(N)],
\label{evol1}
\end{eqnarray}
where $\omega(M)={\rm exp}[\frac{-1}{T(M/N)}]$.
For $\omega(M)\equiv 1$ the above equation are equivalent to the ones of the 
Ehrenfest model~\cite{EHREN}.
Supplementing the above equations with initial conditions one can solve 
them numerically~\cite{nonlinear}.
Examples of such solutions in a long-time limit are shown in Fig.~\ref{config}.
%%%%%%%%%%%%%%%%%%%%%%%%%%%%%%%%%%%%%%%%%%%
%%%%%%%%%%%%%%%%%%%%%%%%%%%%%%%%%%%%%%%%%%%
As expected, one finds a single- and double-peak distribution in regions I 
and II, respectively.
The width of these peaks decreases with the number of particles $N$.
The behaviour of the probability distribution in regions III and IV is however 
more complicated.
Although for small $N$ it has three-peak structure, the shape changes in the 
limit $N\rightarrow \infty$.
The final distribution depends whether the parameters $\Delta$ and $T_0$ 
are in region III or IV.
In region III, for increasing $N$, the central peak diminishes and in the
limit $N\rightarrow \infty$ we obtain a two-peak distribution.
On the other hand, in region IV only the central peak survives in the 
thermodynamic limit.
As we will show, a line which separates these two regions can be interpreted 
as a line of discontinuous transitions.
Before we approximately locate this line, let us study 
fluctuations of the order parameter.
Since we know the probability distribution, these fluctuations can be 
calculated numerically, without making additional assumptions~\cite{EGGERS}.
Using the variance of the order parameter we define the susceptibility $\kappa$
as:
\begin{equation}
\kappa=N<(\epsilon-<\epsilon>)^2>=\frac{1}{N}\{\sum_{i=0}^N i^2p(i,\infty)-
[\sum_{i=0}^N ip(i,\infty)]^2\},
\label{susc}
\end{equation}
where $p(..,\infty)$ denotes the steady-state probabilities obtained from 
eqs.~(\ref{evol1}).
Susceptibility is known to diverge at the continuous phase transition.
Our calculations for $\Delta=0.5$ confirm such a behaviour (Fig.~\ref{susc}) 
For $T_0>0.25$ the inverse susceptibility $\kappa^{-1}$ decays linearly which 
implies $\gamma=1$ (mean-field value).
For $T_0<0.25$ (asymmetric phase) the behaviour is less clear but we also 
expect a linear decay of $\kappa^{-1}$.

To check our steady-state and master-equation calculations we performed Monte
Carlo simulations which are also presented in some cases.
The implementation of this method for the present model is rather 
straightforward.
%%%%%%%%%%%%%%%%%%%%%%%%%%%%%%%%%%%%%%%
\begin{figure}
\centerline{\epsfxsize=8cm 
\epsfbox{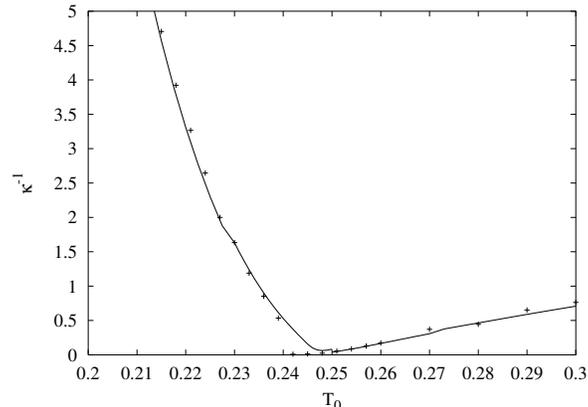}}
%\figspace
\caption{
The inverse of the susceptibility $\kappa^{-1}$ as a function of $T_0$
for $\Delta=0.5$.
Solid line corresponds to the numerical solution of eq.~(\ref{order}) 
for $N=5000$.
Monte Carlo simulations were made for $N=5000$ (+).
In the asymmetric phase ($T_0<0.25$) to calculate $\kappa$ we used only a 
half of the probability distribution.
}
\label{fsusc}
\end{figure}
%%%%%%%%%%%%%%%%%%%%%%%%%%%%%%%%%%%%%%%%%%%
%%%%%%%%%%%%%%%%%%%%%%%%%%%%%%%%%%%%%%%
\begin{figure}
\centerline{\epsfxsize=8cm 
\epsfbox{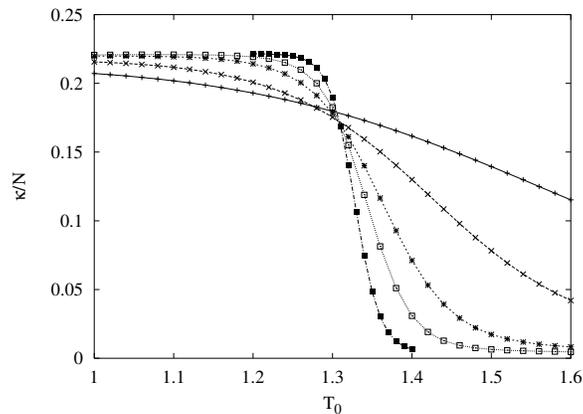}}
%\figspace
\caption{
The scaled susceptibility $\kappa/N$ as a function of $\Delta$
for $T_0=0.2$ obtained from the numerical solution of eq.~(\ref{evol1}) 
for $N=50$, 100, 200, 300, and 500.
}
\label{location}
\end{figure}
%%%%%%%%%%%%%%%%%%%%%%%%%%%%%%%%%%%%%%%%%%%%%%%%%%%%%%%%%%%%%
Let us notice that the behaviour of the probability distribution as seen in
Fig.~\ref{config} bears some similarity to equilibrium systems.
In particular, it is known that the simultaneous existence of peaks of both
phases is an indicator of the first-order phase transitions~\cite{BINDER}.
Although our model is defined only using dynamical rules and thus has no 
free energy (at least as defined for equilibrium systems) it does exhibit some 
other features typical to equilibrium systems.
To precisely locate the transition point we calculated the scaled susceptibility
$\kappa/N$.
We expect that when only a single peak survives $\kappa/N$ vanishes, while it
remains positive in the double-peak region.
For $T_0=0.2$ the obtained results (Fig.~\ref{location}) locate the transition
around $\Delta=1.31$.
As we will see in the next section there is also a dynamical indicator of this
transition.
%%%%%%%%%%%%%%%%%%%%%%%%%%%%%%%%%%%%%%%%%%%%%%%%%%%%%%%%%%%%%%%%%%%%%%%%%%%%%
\section{Metastability and the characteristic time}
Analysis of the probability distribution for our dynamical model which was 
done in the previous section shows some similarity to 
equilibrium discontinuous transitions.
It is known that such transitions are usually accompanied by metastability.
In the present section we examine such effects in our model. 

\begin{figure}
\centerline{\epsfxsize=9cm 
\epsfbox{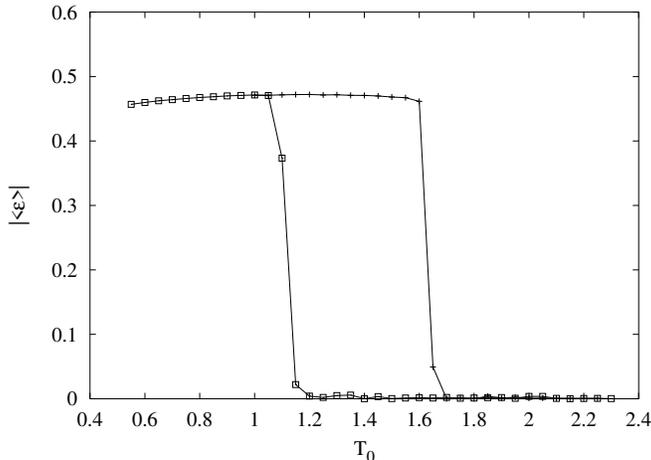}}
%\figspace
\caption{
The absolute value of the order parameter $|<\epsilon>|$ as a function of 
$\Delta$ for $T_0=0.2$ calculated using Monte Carlo simulations for $N=2000$.
Two runs were made with $\Delta$ increasing (+) and decreasing  ($\Box$), 
respectively.
Due to strong metastability no sign of the transition exists for $\Delta=1.31$.
}
\label{hister}
\end{figure}

The most transparent indication of metastability is hysteresis.
That our model exhibits such a behaviour is clearly seen in Fig.~\ref{hister}.
One can see, that upon increasing $\Delta$, the system prepared in the asymmetric 
state remains in this state up to $\Delta\sim 1.7$ which is close to the limits
of stability of this state (see Fig.~\ref{diagram}).
On the other hand, when $\Delta$ decreases the symmetric state persists up to
$\Delta\sim 1.1$ which is again close to the limits of stability of 
symmetric phase.
Simulations show that the range of the hysteresis only slightly depends on the 
system size and the simulation time.
For large system sizes the model will persist in the initial state
until the limits of stability of this state are reached, as calculated from the 
steady-state equation (\ref{flux1}).
Such a behaviour indicates a very strong metastability.
As we will show below, in regions III and IV of the phase diagram certain 
characteristic times diverge exponentially with the system size which 
corresponds to the broken ergodicity (i.e., dynamical phase space of the model
is decomposed into disconnected sectors).
Let us notice that in (short-ranged) equilibrium systems metastability has 
always a finite lifetime and longer simulations or larger system size 
shrink the hysteresis.
%%%%%%%%%%%%%%%%%%%%%%%%%%%%%%%%%%%%%%%%%%%
%%%%%%%%%%%%%%%%%%%%%%%%%%%%%%%%%%%%%%%%%%%
To examine the metastable properties of our model further we will calculate
certain characteristic times.
It is known that for urn models some of these quantities can be calculated
as the so-called first-passage problem.
We will use a similar approach.
More specifically, let us consider a configuration where there are $M$ 
and $N-M$ balls in urn A and B, respectively ($N$-even).
Now, we define an average time $\tau(M,N-M)$ needed for such a configuration to
reach a symmetric configuration ($M=N/2$).
From the dynamical rules of our model one finds that $\tau(M,N-M)$'s obey the 
following $\frac{N}{2}$ linear equations (to simplify notation we omit the 
second argument in $\tau$'s):
\begin{eqnarray}
\tau(N)=&&\omega(N)[\tau(N-1)+1]+[1-\omega(N)][\tau(N)+1]\nonumber\\
\tau(N-1)=&&\frac{N-1}{N}\omega(N-1)[\tau(N-2)+1]+\frac{1}{N}\omega(1)[\tau(N)+1]
+[1-\frac{N-1}{N}\omega(N-1)-\frac{1}{N}\omega(1)][\tau(N-1)+1]\nonumber\\
\vdots\nonumber\\
\tau(M)=&&\frac{M}{N}\omega(M)[\tau(M-1)+1]+\frac{N-M}{N}\omega(N-M)[\tau(M+1)+1]\nonumber\\
&&+[1-\frac{M}{N}\omega(M)-\frac{N-M}{N}\omega(N-M)][\tau(M+1)+1]\nonumber\\
\vdots\nonumber\\
\tau(0.5N+1)=&&(0.5+\frac{1}{N})\omega(0.5N+1)+
(0.5-\frac{1}{N})\omega(0.5N-1)[\tau(0.5N+2)+1]+\nonumber\\
&&[1-(0.5+\frac{1}{N})\omega(0.5N+1)-(0.5-\frac{1}{N})\omega(0.5N-1)]
[\tau(0.5N+1)+1]
\label{taurec}
\end{eqnarray}
Similar equations can be written for the Ehrenfest model~\cite{LIP}.

We were unable to write an explicit solution of the above set of equations.
However, its tridiagonal structure greatly simplifies its numerical solution.
In particular the so-called Gaussian elimination can be used to calculate 
$\tau$'s recursively~\cite{SEDGEWICK}.
Since implementation of this algorithm is straightforward we present only final
results.
Our unit of time corresponds to a single on average update per ball, i.e., we
divide the time $\tau$ calculated from eqs.~(\ref{taurec}) by $N$.

\begin{figure}
\centerline{\epsfxsize=9cm 
\epsfbox{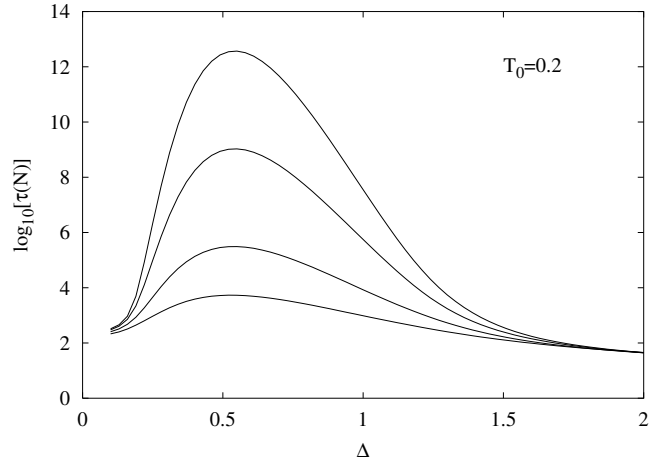}}
%\figspace
\caption{
The characteristic time $\tau(N)$ as a function of $\Delta$ for
(from bottom) $N=50$, 100, 200, and 300.
}
\label{tauex}
\end{figure}

\begin{figure}
\centerline{\epsfxsize=9cm 
\epsfbox{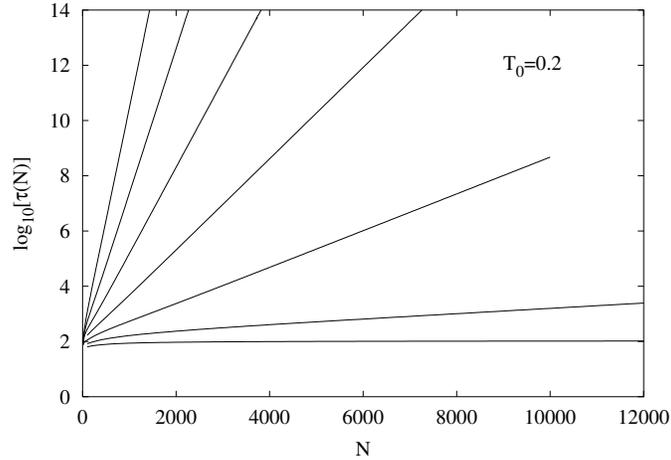}}
%\figspace
\caption{
The characteristic time $\tau(N)$ as a function of $N$ for $T_0=0.2$ and
(from left) $\Delta=1.2$, 1.3, 1.4, 1.5, 1.6, 1.7, and 1.8.
}
\label{taulog}
\end{figure}

\begin{figure}
\centerline{\epsfxsize=9cm 
\epsfbox{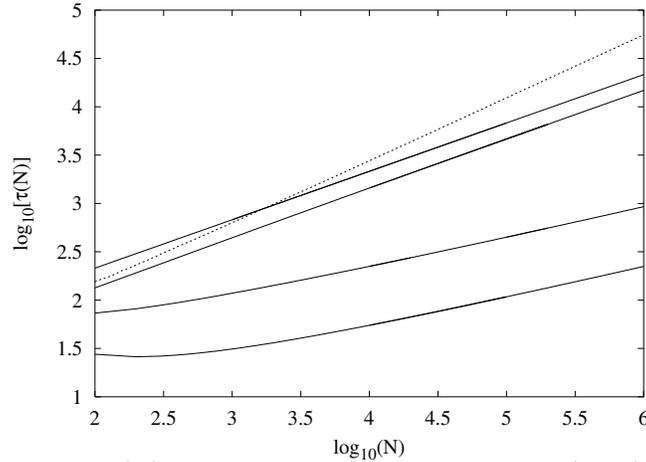}}
%\figspace
\caption{
The characteristic time $\tau(N)$ as a function of $N$  calculated at 
($\Delta,T_0$) equal to (solid lines from top): (0.25,$\sqrt{0.25/2}-0.25/2$), (0.5,0.25),
(1.7440675,0.2), and (6.775493\ldots,0.15)
For two upper lines the points ($\Delta,T_0$) are on the critical line 
(\ref{critical}), while for bottom ones on the lines of the limits of 
stability of the asymmetric solution.
The dotted line corresponds to the tricritical point.
}
\label{taucrit}
\end{figure}

\begin{figure}
\centerline{\epsfxsize=9cm 
\epsfbox{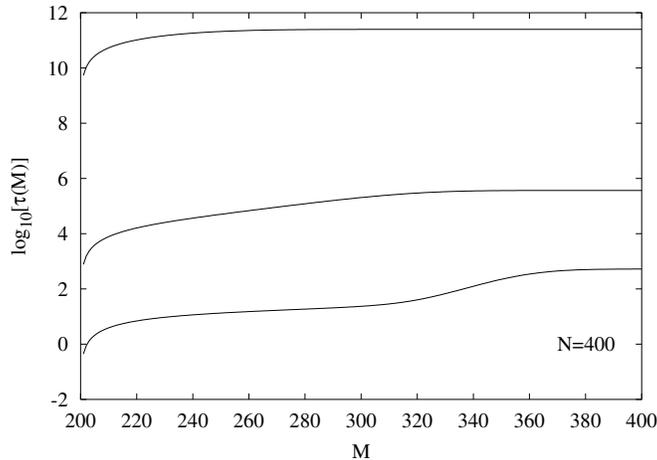}}
%\figspace
\caption{
The distribution of characteristic times $\tau(M)$ as a function of $M$  
calculated $T_0=0.2$ and (from top) $\Delta=0.9$, 1.2, 1.5.
}
\label{taudist}
\end{figure}

\begin{figure}
\centerline{\epsfxsize=9cm 
\epsfbox{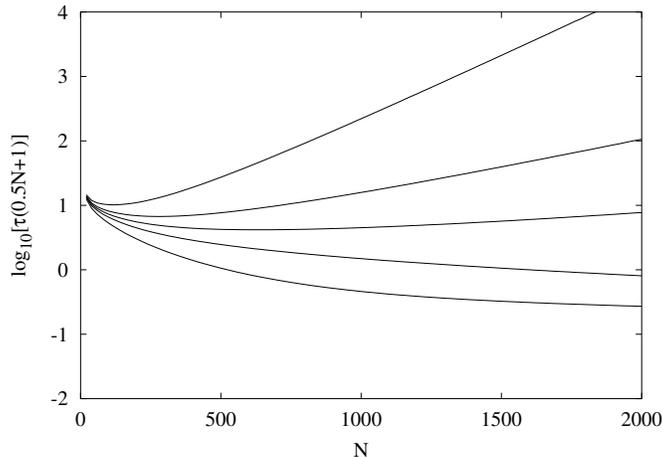}}
%\figspace
\caption{
The characteristic time $\tau(0.5N+1)$ as a function of $N$  
calculated $T_0=0.2$ and (from top) $\Delta=1.27$, 1.29, 1.30, 1.31, and 1.33.
}
\label{tau1size}
\end{figure}

In Fig.~\ref{tauex} we present the $\Delta$-dependence for $\tau(N)$.
This is the lifetime of totally asymmetric solution.
Its rapid increase (as a function of $N$) in certain range of $\Delta$ 
indicates the stability of the asymmetric solution.
%%%%%%%%%%%%%%%%%%%%%%%%%%%%%%%%%%%%%%%%%%%
%%%%%%%%%%%%%%%%%%%%%%%%%%%%%%%%%%%%%%%%%%%
%%%%%%%%%%%%%%%%%%%%%%%%%%%%%%%%%%%%%%%%%%%
\begin{figure}
\centerline{\epsfxsize=9cm 
\epsfbox{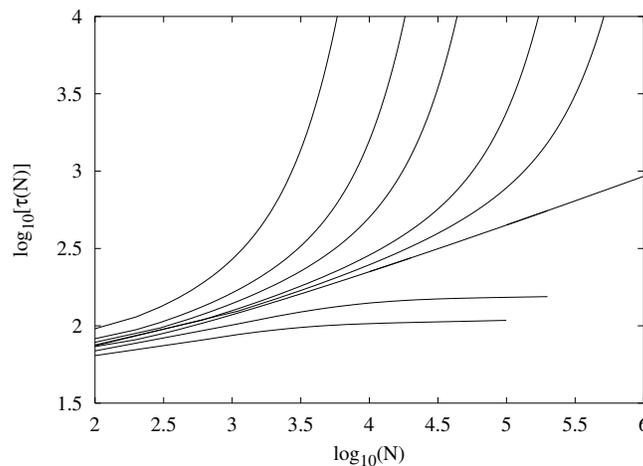}}
%\figspace
\caption{
The characteristic time $\tau(N)$ as a function of $N$ for $T_0=0.2$ and
(from left) $\Delta=1.65$, 1.7, 1.72, 1.735, 1.74, 
$\Delta_c(0.2)=1.7440675\ldots$, 1.77, and 1.8.
}
\label{tau02}
\end{figure}
%%%%%%%%%%%%%%%%%%%%%%%%%%%%%%%%%%%%%%%%%%%
To provide more details on the stability of this solution we have to
look at the size dependence of the characteristic time and the results are 
shown in Fig.~\ref{tau02}-\ref{taulog}.
One can see (Fig.~\ref{tau02}) that the limits of stability of asymmetric 
solution, which for $T_0=0.2$
equals $\Delta=\Delta_c(0.2)=1.7440675\ldots$, separates two regimes.
For $\Delta>\Delta_c(0.2)$ $\tau(N)$ approa ches a finite value for increasing
$N$, while it diverges for $\Delta<\Delta_c(0.2)$.
From Fig.~\ref{taulog} one can conclude that for $\Delta<\Delta_c(0.2)$,
$\tau(N)$ diverges exponentially fast with $N$.
It confirms our previous observations, based on Monte Carlo simulations, that
a given phase persists up to the limits of its stability.
%%%%%%%%%%%%%%%%%%%%%%%%%%%%%%%%%%%%%%%%%%%
%%%%%%%%%%%%%%%%%%%%%%%%%%%%%%%%%%%%%%%%%%%
Form Fig.~\ref{taucrit} it follows that at the limits of stability $\tau(N)$
increases but as a power of $N$.
From the slope of these data we estimate that both for $T_0=0.2$ and 0.15 $\tau\sim N^{1/3}$.
A different exponent governs the increase of $\tau(N)$ along the critical line 
(\ref{critical})~\cite{comment1}.
In this case we find $\tau(N)\sim N^{1/2}$, while at the tricritical point 
$N^{2/3}$ increase is observed (see Fig.~\ref{taucrit}).
It would be desirable to explain such simple power laws using analytical 
arguments.
%%%%%%%%%%%%%%%%%%%%%%%%%%%%%%%%%%%%%%%%%%%
%%%%%%%%%%%%%%%%%%%%%%%%%%%%%%%%%%%%%%%%%%%
The above results concern the behaviour of $\tau(N)$.
We have shown that this quantity can be used to locate the critical line or the 
limits of stability of the asymmetric  solution.
However, this quantity does not provide any indication of the discontinuous phase
transition which we located between regions III and IV on Fig.~\ref{diagram}.
To have a dynamical indication of this transition we have to look for other 
characteristic times.
First, let us notice (Fig.~\ref{taudist}) that in region II 
the distribution of $\tau$'s as function of $M$ is rather flat which means 
that there is basically a single time scale in the model.
In regions III and IV, however, one can see a more pronounced variability 
of $\tau(M)$.
In Fig.~\ref{tau1size} we show that $\tau(0.5N+1)$ (which is the shortest 
of our $\tau$'s) can be used to locate a discontinuous phase transition.
Indeed, this quantity increases exponentially fast with $N$ for $\Delta<1.31$ 
and remains bounded for larger values of $\Delta$.
The change of behaviour of $\tau(0.5N+1)$ coincides, within our numerical
accuracy, with the change of the probability  distribution as shown in 
Fig.~\ref{location}
%%%%%%%%%%%%%%%%%%%%%%%%%%%%%%%%%%%%%%%%%%%
%%%%%%%%%%%%%%%%%%%%%%%%%%%%%%%%%%%%%%%%%%%
%%%%%%%%%%%%%%%%%%%%%%%%%%%%%%%%%%%%%%%%%%%
%%%%%%%%%%%%%%%%%%%%%%%%%%%%%%%%%%%%%%%%%%%
Let us also notice that the time scale reached in the numerical solution of
eqs.~(\ref{taurec})
($\sim 10^{14}$) is much larger than what is accessible using Monte Carlo
simulations.
The main limitation in our calculations is finite accuracy of the 
computations.
Using longer representations of real numbers (we used the FORTRAN real*8 type )
one can study much larger time scales.
%%%%%%%%%%%%%%%%%%%%%%%%%%%%%%%%%%%%%%%%%%%%%%%%%%%%%%%%%%%%%%%%%%
\section{conclusions}
In the present paper we examined an urn model which undergoes a symmetry 
breaking transition.
Our work was motivated by recent experiments on the separation of shaken sand.
The proposed model exhibits a rich behaviour (continuous/discontinuous 
transitions, a tricritical point, metastability) and its properties could be 
reliably determined.
There are experimental indications that in the three-compartment case the 
separation of shaken sand is a discontinuous transition~\cite{MEER}.
However, due to strong metastability only the hysteretic behaviour is observed,
similarly as in our simulations in Fig.~\ref{hister}.
Our work suggests, that there are some indicators (probability distributions,
characteristic times) which could be used to locate  a discontinuous transition
in such systems.
It would be interesting to look for such indicators in experimental systems.

As an extension of our work, one can examine models with a different relation 
between the effective temperature and the number of particles or models with
larger number of compartments.
Even further extensions are related with a much different interpretation of our 
model.
One can imagine that two compartments A and B are two groups of people 
(e.g., supporters of certain presidential candidates).
It seems reasonable to assume that under certain conditions a larger number 
of people in a given group will increase attractive bonds within this group
(energy).
On the other hand it will also increase a probability that a certain member 
will leave the group (entropy).
Competition of this two effects is likely to produce a similar symmetry 
breaking transition to the one described in the present paper.
However, since people are known to create various local connections 
(e.g. within a group) it would be desirable to take such effects into 
account too.
In our model particles are chosen randomly and such correlations are omitted.
%%%%%%%%%%%%%%%%%%%%%%%%%%%%%%%%%%%%%%%%%%%
%%%%%%%%%%%%%%%%%%%%%%%%%%%%%%%%%%%%%%%%%%%%%%%%%%%%%%%%%%%%%%%%%%%%
%%%%%
\acknowledgements
This work was partially supported by the Swiss National Science Foundation
and the project OFES 00-0578 "COSYC OF SENS".
%%%%%%%%%%%%%%%%%%%%%%%%%%%%%%%%%%%%%%%%%%%%%%%%%%%%%%%%%%%%%%%%%%%%%%%%%%

%%%%%%%%%%%%%%%%%%%%%%%%%%%%%%%%%%%%%%%%%%%%%%%%%%%%%%%%%%%%%%%%%%%%
%%
%\end{multicols}

\begin{references}
\bibitem{even} One can easily notice that even-order coefficients in the 
expansion of both sides of eq.~(\ref{flux1}) are the same and thus cancel 
each other.
\bibitem{nonlinear} Analytical approach to evolution equations~(\ref{evol1})
seems to be very difficult since the variable $M$  appears in an exponential
factor $\omega(M)$.
\bibitem{comment1} Actually, only a right-hand (respective to the tricritical 
point) portion of the curve 
(\ref{critical}) can be called a critical line and only on this part the 
scaling $\tau(N)\sim N^{1/2}$ is observed.
The left-hand side is only the limit of the stability of the symmetric 
solution.
Since the asymmetric solution is stable on this part, $\tau(N)$ diverges
exponentially with $N$.
\end{references}
\end{document}